# Mobile Néel skyrmions at room temperature: status and future

Wanjun Jiang[1,†], Wei Zhang[1], Guoqiang Yu[2], M. Benjamin Jungfleisch[1], Pramey Upadhyaya[2,3], Hamoud Somaily[1,4], John E. Pearson[1], Yaroslav Tserkovnyak[3], Kang L. Wang[2], Olle Heinonen[1], Suzanne G. E. te Velthuis[1,†], and Axel Hoffmann[1,†]

[1]Materials Science Division, Argonne National Laboratory, Lemont, Illinois, USA, 60439

[2]Department of Electrical Engineering, University of California, Los Angeles, California USA, 90095

[3]Department of Physics and Astronomy, University of California, Los Angeles, California, USA, 90095

[4]Department of Physics, Northern Illinois University, DeKalb, IL 60115

[†]To whom correspondences should be addressed.

E-mail:

jiangw@anl.gov

tevelthuis@anl.gov

hoffmann@anl.gov

**Abstract**

Magnetic skyrmions are topologically protected spin textures that exhibit many fascinating features. As compared to the well-studied cryogenic Bloch skyrmions in bulk materials, we focus on the room-temperature Néel skyrmions in thin-film systems with an interfacial broken inversion symmetry in this article. Specifically, we show the stabilization, the creation, and the implementation of Néel skyrmions that are enabled by the electrical current-induced spin-orbit torques. Towards the nanoscale Néel skyrmions, we further discuss the challenges from both material optimization and imaging characterization perspectives.

**Introduction**

After their original observation in MnSi single crystals[1-3], magnetic (Bloch) skyrmions have stimulated tremendous research efforts in the field of spintronics[4-9]. On the one hand, thanks to the well-defined spin topology in real space, magnetic skyrmions enable many intriguing quantum-mechanical phenomena to be observed including: emergent electromagnetic dynamics[10], effective magnetic monopole[11], topological/skyrmion Hall effects[7, 9, 12]. On the other hand, due to their topological properties, magnetic skyrmions can behave as meta-stable quasi-particles and thus have been envisioned as information carriers for ultra-low power non-volatile spintronics[4, 5, 13].

Although the investigation of magnetic skyrmions was pioneered using bulk materials with chiral exchange interactions due to crystal symmetries lacking inversion, there are so far only very few materials that enable the stabilization of Bloch skyrmions above room temperature[14]. In contrast, recent progress on interfacial nanomagnetism has significantly extended the paradigm of magnetic skyrmions to readily accessible materials system – heavy metal/ultra-thin ferromagnet/insulator (HM/FM/I) hetero-structures with perpendicular magnetic anisotropy[7, 13, 15-22]. The interfacial symmetry breaking in HM/F/I hetero-structures introduces a chiral interfacial Dzyaloshinskii-Moriya interactions (DMI) between the neighbouring atomic spins $S_1$ and $S_2$ that can be written as $-D_{dmi} \cdot (S_1 \times S_2)$, where the DMI vector $D_{dmi}$ lies in the film plane acting as an equivalent (spatial) in-plane field and stabilizes Néel (hedgehog) skyrmions/domain walls with a fixed chirality[18, 21, 23-26]. In these interfacial symmetry-breaking systems, the stabilization, generation, and manipulation of hedgehog skyrmions by electric currents have been demonstrated at room temperature[21, 27, 28].

Beyond the realization of room-temperature magnetic skyrmions, this novel material system provides many unique advantages. From a material synthesis perspective: these thin films and their multilayers can be readily produced by using a magnetron sputtering technique onto $SiO_2$ substrates. From an application perspective, as a result of the strong spin-orbit interaction of the involved heavy metals (typically Ta, Pt and W)[13, 29, 30], the electrical current induced spin-orbit torques from the spin Hall effects provide very energetically efficient avenues for electrically generating, manipulating, and more importantly, implementing magnetic skyrmions at room temperature.

**Experiment and Discussion**

The feasibility of generating mobile magnetic skyrmions at room temperature on demand was demonstrated by using patterned heterostructures[19]. A trilayer of Ta(5nm)/$Co_{20}Fe_{60}B_{20}$(CoFeB)(1.1nm)/$TaO_x$(3nm) grown at room temperature by a magnetron sputtering technique onto a $SiO_2$ substrate. A polar magneto-optical Kerr effect (MOKE) microscope in a differential mode was utilized for imaging experiments at room temperature. By introducing a geometrical constriction that spatially modulates the distribution of electric currents in the HM/F/I trilayer with an interfacial DMI, we have previously demonstrated that the resultant divergent spin-orbit torques could continuously transform the chiral band domains into hedgehog skyrmions with a fixed (left-handed) chirality, as shown in Figs.1 (a) – (g)[13, 21]. This dynamical conversion is enabled by the increased surface tension of the expanding chiral band domain around the constriction, as a result of the divergent spin Hall effective forces/fields. These artificially magnetic skyrmions, once created, move along the electron current direction, without noticeable deformation within the instrument limit. It should mention here that these dynamical instabilities are similar to Rayleigh-Plateau instabilities in surface-tension dominated fluid dynamics[31]. However, to precisely quantify the underlying physics, further theoretical efforts and numerical simulation studies are required. Nevertheless, the skyrmion spin texture can be inferred from the current driven motion by comparing the response to a charge current of topologically trivial magnetic bubble domains. For this magnetic bubble domains were stabilized in an in-plane magnetic field sufficiently large to orient the in-plane magnetization of the domain walls into a single direction. Instead of freely moving along the currents as observed for the artificially synthesized skyrmions, these topological trivial bubbles either shrink or expand, depending on the polarity of currents[19].

Based on these results, interesting questions that arise are: Does the width of the geometrical constrictions matter? Once the geometrical constriction is wider than the width of band domain, band domains move smoothly through the constriction, without suffering from edge repulsion. The dynamical conversion from band domains into magnetic skyrmions occurs as expected, as shown in the Figs. 2 (a) – (b) for a constriction of width 8 μm. On the other hand, when the geometrical constriction (1 μm) gets narrower than the width of the band domains ($\approx$ 1 μm), we have also observed the generation of magnetic skyrmions, see Figs. 2 (c)–(d). While the exact magnetic domain behavior in the constriction cannot be resolved optically, it is clear that the charge current inhomogeneity after the construction still can generate new skyrmions. Furthermore, with decreasing constriction width the current inhomogeneity increases and thus smaller currents are sufficient, as can be seen in Fig. 2 (e).

As experimentally demonstrated, a geometrical constriction facilitates the formation of magnetic skyrmions, but it also impedes a systematic investigation of spin-orbit torques induced motion of skyrmions, as well as a precise control of individual skyrmion. This is due to the efficient production of hedgehog skyrmions through the constriction under the higher current densities, which make it difficult to distinguish the behavior of individual skyrmions. At the same time, the maximum current density necessary to explore the high velocity regime, is limited due to the width of the constriction. To overcome this dilemma, we designed a four-terminal skyrmionic device, in which the skyrmion generation line (vertical) and skyrmion motion line (horizontal) are separated, as shown in Fig. 3. After passing a current pulse for skyrmion generation ($I_{gen+}$ to $I_{gen-}$) through the constriction of width 4 μm and of length 15 μm, synthetic skyrmions are created in the part with a divergent current distribution. Subsequently, another current pulse for skyrmion motion ($I_{mot+}$ to $I_{mot-}$) is applied to the skyrmion motion line (of width 8 μm) to "load" synthetic skyrmions into this wire. By narrowing the width of skyrmion motion line, it is possible to isolate and subsequently study the current driven motion of a single skyrmion driven by the spin-orbit torques, as shown in Figs.4 (a) – (i). It is noted that, due to edge repulsion, the skyrmion Hall effect (where skyrmion moves perpendicular to the current direction) is not resolved experimentally.

From the images in Fig, 4, it can be see that the skyrmions are displaced by about 10 μm for each 10 μs current pulse of 0.5 MA/cm$^2$. This corresponds to a velocity of 1 m/s, which is comparable to current driven motion of chiral domain walls at similar current densities. Furthermore, the displacement is identically reproduced for subsequent current pulses as long as the skyrmion is not in the vicinity of the additional voltage contacts. This indicates that for these higher current densities random pinning is inconsequential. However, when the skyrmion is close to the additional voltage contacts of the wire, the displacement is somewhat reduced. This could either be due to current spreading into the voltage contacts, which effectively reduce the current density, or due to dipolar interactions of the skyrmions with the magnetic material forming the voltage contacts.

**Perspectives**

While tremendous progresses have been made on the emerging field of magnetic skyrmions in thin film heterostructures, there are many challenges that still need to be addressed. In order to make magnetic skyrmions truly attractive for ultra-high density data storage, and to probe at room temperature the emergent new physics due to their topology, such as topological or skyrmion Hall effects, the size of the magnetic skyrmions has to be significantly reduced[21, 32, 33]. The size of magnetic skyrmions shown here is large (≈ 1 μm in diameter) due to its relatively weak DMI (≈ 0.5±0.1 mJ/m²) at the Ta/CoFeB interface, as compared to other material systems such as Pt/Co (≈ 1.3±0.1 mJ/m²). It should also be noted that the insulator layer in our Ta/CoFeB/TaO$_x$ heterostructures contributes to the perpendicular magnetic anisotropy and interfacial asymmetry, but its role for interfacial DMI is unclear. By utilizing different signs/strengths of spin Hall angle from various transition metals, Woo *et al.*, has successfully observed magnetic skyrmions of a few hundreds nanometer in diameter at room temperature in a Pt/Co/Ta multilayer[27]. Interfacing thin ferromagnets between two different transition heavy metals may be advantageous for minimizing the size of skyrmions, and for boosting the motion of skyrmions, since materials with opposite sign of spin Hall angles [*e.g.*, Pt , $\theta_{sh}$≈ +10%, and Ta, $\theta_{sh}$≈ -20%] at opposite interfaces can provide complementary spin-orbit torques. By taking the advantage of opposite DMIs at two interfaces[33], Moreau-Luchaire *et al.*, has successfully observed skyrmions of few tens of nanometer in a Pt/Co/Ir multilayer[34]. Properly designing the DMI in these heterostructures may be more complicated[34], since the magnitude of the DMI might depend strongly on the details of local crystallographical microstructure.

Smaller skyrmions also provide additional challenges for their characterization. Upon approaching few tens of nanometer Néel skyrmions at room temperature, one has to access novel spin sensitive imaging techniques to reveal the associated chirality. Lorentz transmission electron microscopy (L-TEM) has been a powerful tool to study the Bloch-type skyrmions in B20 compounds[3]. However, the phase shift for a given Néel skyrmion/domain wall is zero in the Lorentz mode, which results in the absence of magnetic contrast. On the other hand, it is known that the spin-polarized low-energy electron microscopy (SPLEEM) [15, 16, 35] and spin-polarized scanning tunnelling microscopy (SP-STM)[33] are capable of quantifying all 3-dimensional components of an arbitrary spin texture. However, the sample preparation for these techniques is non-trivial and often these techniques do not work for the ex-situ patterned devices. A more suitable approach for ex-situ patterned devices may be a nitrogen vacancy (NV) center diamond microscope, is expected to be suitable for quantifying the 3 dimensional arbitrary spin textures with a sub-nm resolution[36, 37]. This technique can thus in principle be used to map out the spin topology, the dynamics, and the deformation of nanometer skyrmions, driven by both spin-orbit torques and magnetic fields, but so far this has not yet been adopted to the investigation of Néel skyrmions.

**Conclusions**

In summary, by harvesting the interfacial Dzyaloshinskii-Moriya interactions and the current induced spin-orbit torques in the heavy metal/ultra-thin ferromagnet/insulator heterostructures, Néel skyrmion might be more advantageous as compared to Bloch skyrmion in bulk crystals, including the efficient creation, manipulation, and implementation. It could enable not only the realization of functional skyrmionic devices, but also the observation of intriguing topological transport phenomena possibly at room temperature.

**Acknowledgements**

Work carried out at the Argonne National Laboratory, including patterned device fabrication, magneto-optic imaging and data analysis, was supported by the U.S. Department of Energy, Office of Science, Materials Science and Engineering Division. Lithography was carried out at the Center for Nanoscale Materials, which is supported by the DOE, Office of Science, Basic Energy Science under Contract No. DE-AC02-06CH11357. Work performed at UCLA, including growth of Ta/CoFeB/TaO$_x$ trilayers, was partially supported by the NSF Nanosystems Engineering Research Center for Translational Applications of Nanoscale Multiferroic Systems (TANMS).

**Figure captions:**

Figure 1. (a)-(d) Schematics of magnetic skyrmion formation in constricted wires (a) Spatially varying current distribution around a geometrical constriction, given by red dashed arrows. The solid arrows indicate the in-plane magnetization directions of the domain walls between up- and down-magnetic domains. (b) The vertical component of current. (c) The expansion of band domains due to the force from spin orbit torques $F_{sh}^{y}$. $F_{res}$ is a restoring force due to the surface tension of the domain wall. (d) The final formation of synthesized Néel skyrmions. (e) Domain structures acquired by using a polar MOKE microscope imaging both sides of the device at a perpendicular magnetic field of $B_\perp$ = -0.5 mT. The width of the constriction is 3 μm. (e) and (f) After passing a pulse current of $j_e$ = +5×10$^5$ A/cm$^2$ of duration 1 s, the left side of the device shows elongated band domains, and the other side shows magnetic skyrmions. Current density was normalized by using the wider part of the device. Reprinted with permission from AAAS.

Figure 2. (a) The formation of synthetic skyrmions as a function of the widths of geometrical constriction. If the constriction is bigger than the width of the band domain, band domains flow smoothly through and produce the synthetic skyrmions at the other side, (a) before current and (b) after applying current for constriction of width 8 μm. On the other hand, we also observe the formation of synthetic skyrmions if the constriction is smaller than the width of the band domain, (c) before current and (d) after applying current for constriction of width 1 μm. The threshold of converting into skyrmions as a function of widths of constriction is summarized in (e).

Figure 3. MOKE image of the experimentally constructed skyrmionic device that enables the investigation of a single magnetic skyrmions moving along a wire.

Figure 4. Motion of a single synthetic skyrmion along a wire enabled by applying a pulse current of amplitude 0.5 MA/cm² and duration 10 µs. By dividing the average displacement with the pulse duration, the average velocity $v$ = 1 m/s can be determined. It is noted that the Hall leads shunt off the current distribution and result in a varying (smaller) displacement.

Figure 1

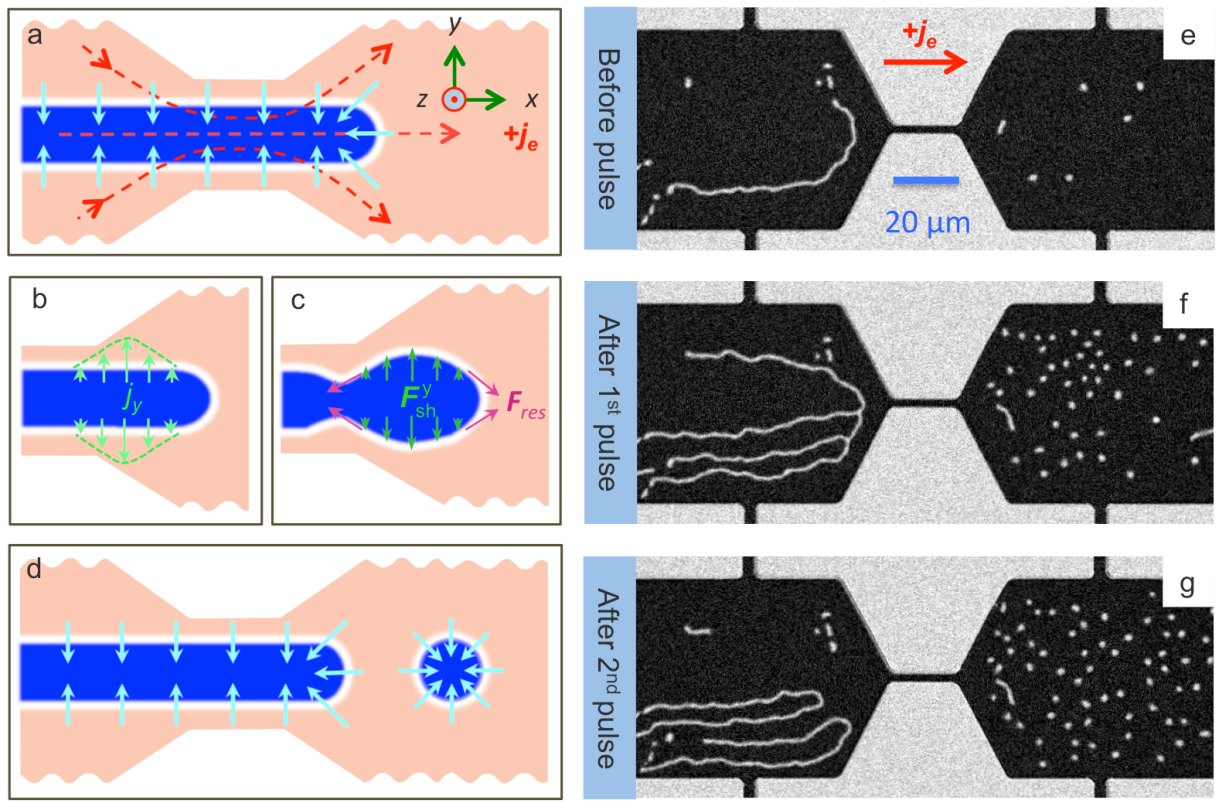

Figure 2

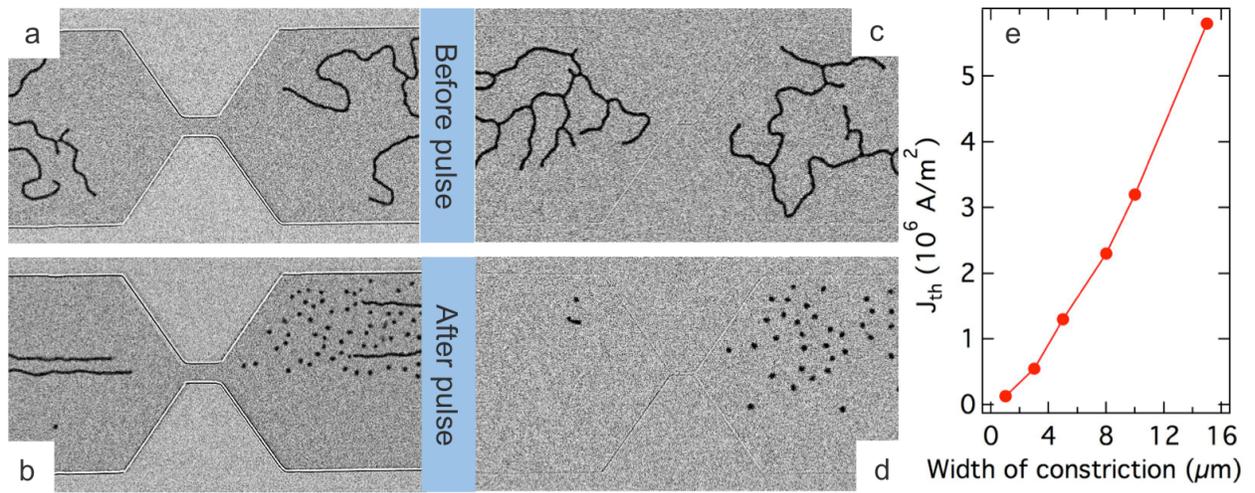

Figure 3

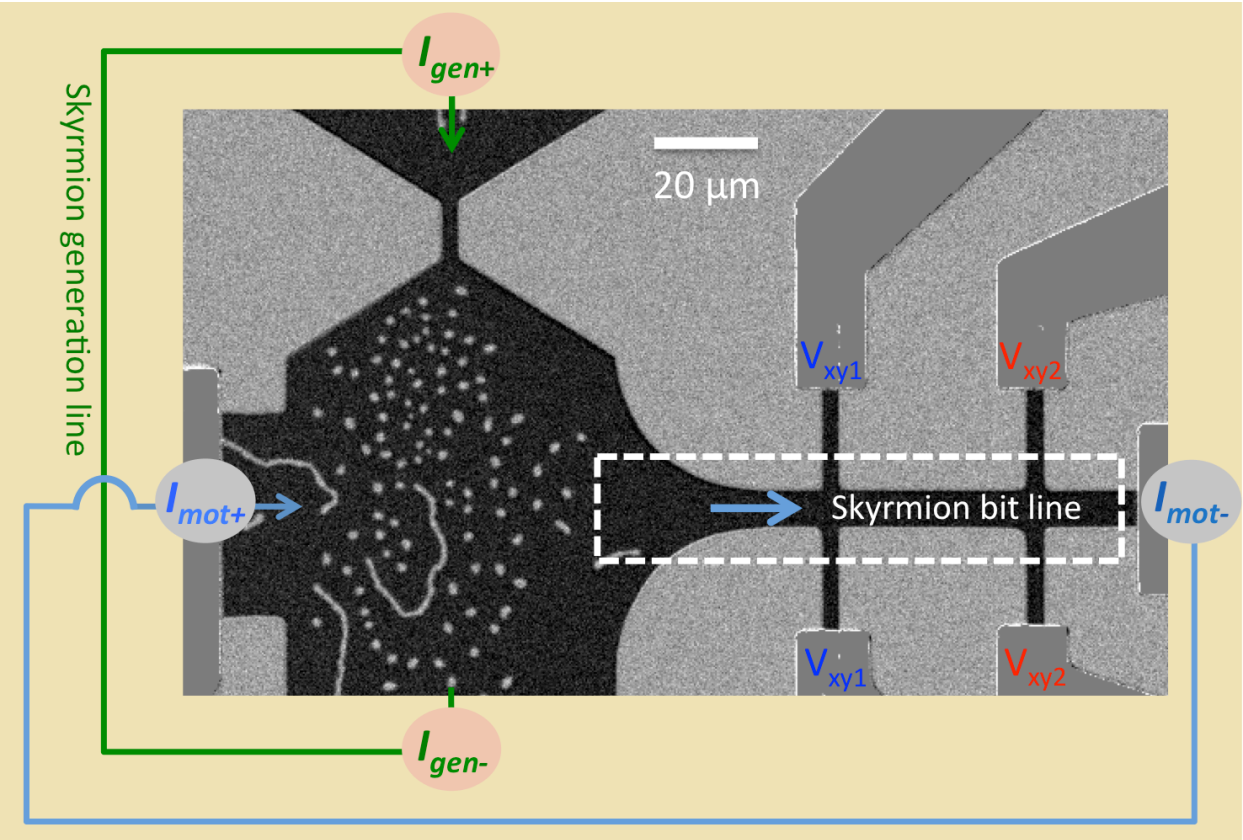

Figure 4

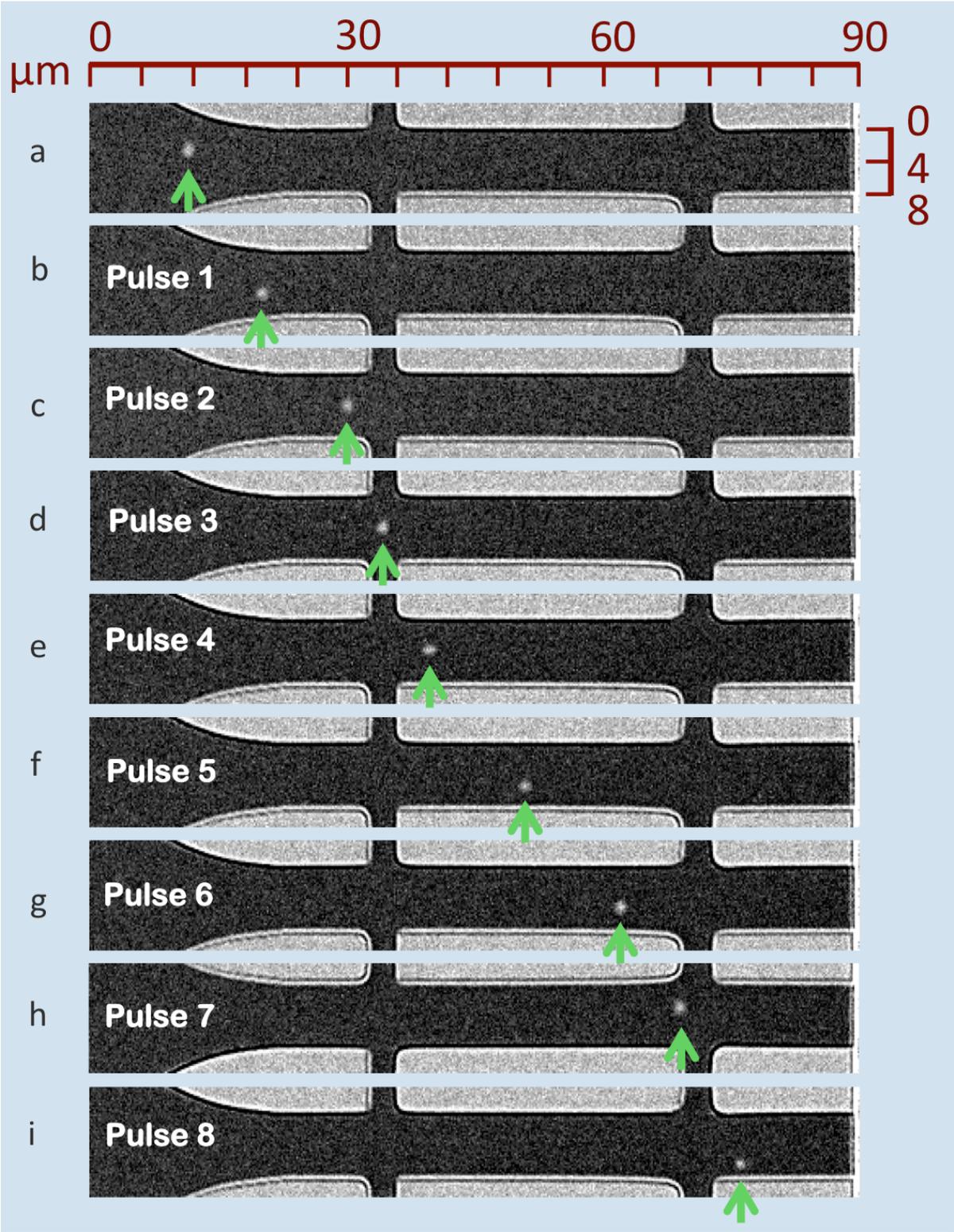